\documentclass[preprint]{aastex}

\def\kms{\ifmmode{\rm km\thinspace s^{-1}}\else km\thinspace s$^{-1}$\fi}
\def\GSC{GSC~01944-02289}

\shorttitle{Wide-field Transit Surveys}

\begin{document}

\title{The Challenge of Wide-Field Transit Surveys: The Case of GSC~01944-02289}

\author{Georgi Mandushev\altaffilmark{1}, Guillermo Torres\altaffilmark{2}, 
David W. Latham\altaffilmark{2}, David Charbonneau\altaffilmark{2,3}, 
Roi Alonso\altaffilmark{4}, Russel J. White\altaffilmark{3}, 
Robert P. Stefanik\altaffilmark{2}, Edward W. Dunham\altaffilmark{1}, 
Timothy M. Brown\altaffilmark{5}, and Francis T. O'Donovan\altaffilmark{3}}

\altaffiltext{1} {Lowell Observatory, 1400 W. Mars Hill Rd., Flagstaff, AZ 
86001, USA; gmand@lowell.edu; Ted.Dunham@lowell.edu}
\altaffiltext{2} {Harvard-Smithsonian Center for Astrophysics, 60 Garden 
St., Cambridge, MA 02138, USA; gtorres@cfa.harvard.edu; dlatham@cfa.harvard.edu;
dcharbonneau@cfa.harvard.edu; rstefanik@cfa.harvard.edu}
\altaffiltext{3} {California Institute of Technology, MC 105-24, 1200 E. 
California Blvd., Pasadena, CA 91125, USA; rjw@astro.caltech.edu; 
ftod@astro.caltech.edu}
\altaffiltext{4} {Instituto de Astrof{\'\i}sica de Canarias, C/ v{\'\i}a 
L\'actea s/n, 38200 La Laguna, Tenerife, Spain; ras@iac.es}
\altaffiltext{5} {High Altitude Observatory, National Center for Atmospheric 
Research, P.O. Box 3000, Boulder, CO 80307, USA; timbrown@hao.ucar.edu}

\begin{abstract}

Wide-field searches for transiting extra-solar giant planets face the 
difficult challenge of separating true transit events from the numerous false 
positives caused by isolated or blended eclipsing binary systems. We describe 
here the investigation of \GSC, a very promising candidate for a transiting 
brown dwarf detected by the Transatlantic Exoplanet Survey (TrES) network. The 
photometry and radial velocity observations suggested that the candidate was an 
object of substellar mass in orbit around an F star. However, careful analysis 
of the spectral line shapes revealed a pattern of variations consistent with the 
presence of another star whose motion produced the asymmetries observed in the 
spectral lines of the brightest star. Detailed simulations of blend models 
composed of an eclipsing binary plus a third star diluting the eclipses were 
compared with the observed light curve and used to derive the properties of the 
three components. Using the predicted stellar parameters we were able to 
identify a second set of spectral lines corresponding to the primary of the 
eclipsing binary and derive its spectroscopic orbit. Our photometric and 
spectroscopic observations are fully consistent with a blend model of a 
hierarchical triple system composed of an eclipsing binary with G0V and M3V 
components in orbit around a slightly evolved F5 dwarf. The rotational 
broadening of the spectral lines of the F5 primary ($v\sin i \approx 34$~\kms) 
and its brightness relative to the eclipsing binary ($\sim 89\%$ of the total 
light) made the discovery of the true nature of the system particularly 
difficult. We believe that this investigation will be helpful to other groups 
pursuing wide-field transit searches as this type of false detection could be 
more common than true transiting planets, and difficult to identify.

\end{abstract}

\keywords{binaries: eclipsing --- line: profiles --- planetary systems --- 
 techniques: photometric --- techniques: radial velocities}

\section{Introduction}

Nearly all of the more than 130 known extrasolar planets\footnote{Extrasolar 
Planets Catalog (http://cfa-www.harvard.edu/planets/catalog.html), 
maintained by Jean Schneider.} have been discovered by means of radial velocity 
observations. Such observations provide no information on the planet's size, 
and, because of the unknown orbital inclination, only a lower limit on the mass 
of the planet can be established. One of the most important results from the 
radial velocity planet detections was that the extrasolar planetary systems 
discovered so far are very different from the solar system, with massive (and 
presumably large) Jupiter-like planets orbiting very close (less than 0.1~AU) to 
their central stars.

Long before the initial discoveries of extrasolar planets around Sun-like stars 
by the Doppler technique \citep{Lath89,Mayo95}, it was recognized that planets 
around other stars could also be detected photometrically if they cross in front 
of their parent stars producing transits \citep{Stru52}. Transit observations 
can provide the missing planet size information and in combination with radial 
velocity data they can be used to derive the planet's density and infer its 
composition. The realization that there could be many extrasolar Jupiter-like 
planets in short-period orbits led to the development of a number of 
ground-based, wide-field photometric searches for transiting 
planets\footnote{Transit Search Programmes 
(http://star-www.st-and.ac.uk/$\sim$kdh1/transits/table.html), maintained by 
Keith Horne.}. In those surveys many thousands of relatively bright stars 
($9 \lesssim R \lesssim 13$) are monitored for periodic shallow drops in the 
star brightness caused by a transiting planet. The expected amplitude of the 
drop and duration of the transit are 1\%--2\% and 2--4 hr, respectively, for a 
Jupiter-sized planet in a 2-7 day orbit around a Sun-like star. Recent advances 
in instrumentation and data analysis allow measurements of the stellar 
brightness with milli-magnitude precision over a period of several weeks or 
months with small, relatively inexpensive and automated telescopes. Indeed, the 
Transatlantic Exoplanet Survey (TrES) network recently announced the first 
successful detection of a transiting planet using a wide-field, small-aperture 
transit survey \citep{Alon04}.

Confirmation of the planetary nature of the transit candidates still requires 
follow-up radial velocity observations with larger telescopes since photometry 
alone cannot distinguish true planetary transits from transit-like events 
produced by other astronomical phenomena \citep{Brow03,Lath03,Char04}. Common 
impostors are eclipsing binaries in which the secondary is much smaller than 
the primary, or systems with components undergoing grazing eclipses. In both 
cases the radial velocity amplitudes are tens of kilometers per second, and 
usually only a few low-precision ($\sim 1$~\kms) radial-velocity measurements 
are enough to reveal the stellar nature of the companions. In other cases more 
subtle analyses of the photometric and spectroscopic data are needed to 
ascertain that the transit candidate is in fact a planet 
\citep[see, e.g.,][]{Kona03,Torr04a,Torr04b}. 

While there is very little published observational data on the rate of false 
alarms in wide-field transit surveys, the statistics in \citet{Dunh04} and the 
preliminary results from the Vulcan survey \citep{Lath03} indicate that the 
frequency of false positives is very high, probably 20 or more times the number 
of true planetary transits. In the case of TrES-1, the only transiting planet 
discovered in such a survey \citep{Alon04}, the ratio of transit candidates to 
true planets was 25:1. Therefore, it is to be expected that most of the 
transit candidates in any photometric survey will be false positives and careful 
follow-up work is crucial in establishing their true nature.

In this paper we describe the case of \GSC, a relatively bright 
($V \simeq 10.4$) star in Cancer that initially appeared to be a very good 
candidate for a transiting brown dwarf. The transit light curves and the radial 
velocity observations were consistent with a $0.03 M_{\sun}$ brown dwarf in a 
$3.35$ day orbit around an F5 star. However, further analysis shows that this is 
not a transiting brown dwarf, but instead a blend of an eclipsing binary and a 
slightly evolved F5 star in a hierachical triple system.

\section{Wide-Field Photometry \label{sec:widefield}}

A $6\degr \times 6\degr$ field centered on $\chi$~Cancri was observed with the 
Lowell Observatory automated Planet Search Survey Telescope (PSST) between 2003 
February 3 and April 26. This telescope is part of TrES, with telescopes at 
Palomar Observatory (USA), Lowell Observatory (USA), and Instituto de 
Astrof{\'\i}sica de Canarias (Spain). The PSST is a 105~mm f/2.8 refractor 
equipped with a $2{\rm K} \times 2{\rm K}$ CCD and is described fully in 
\citet{Dunh04}. Nearly all exposures were taken through a Kron-Cousins $R$\/ 
filter with the exception of several hundred $B$\/ and $V$\/ images on 2003 
March 7 and 8 that were used to estimate stellar colors. A total of 4894 
$R_{\rm C}$\/ exposures were used in the photometric analysis after the 
rejection of images of questionable quality. All $R_{\rm C}$\/ images have an 
exposure time of 90~s. 

Our photometric reduction pipeline is built around the difference image 
analysis (DIA) approach described in \citet{Alar00}. The complete pipeline is 
explained in detail in \citet{Dunh04}; here we outline only the principal steps. 

All images were flat-fielded and overscan- and bias-subtracted using the 
appropriate tasks in the IRAF\footnote{IRAF is distributed by the National 
Optical Astronomy Observatory, which is operated by the Association of 
Universities for Research in Astronomy, Inc., under cooperative agreement with 
the National Science Foundation.} package \citep{Tody93}. One of the images 
taken at low airmass on a moonless photometric night was chosen as the 
reference image for the field, and profile-fitting photometry was carried out 
for that image using Peter Stetson's {\sc daophot~ii}/{\sc allstar} suite 
\citep{Stet87,Stet92}. After deriving the geometric transformations between the 
reference image and all other images, all images were resampled to the pixel 
grid of the reference image. A master frame for the field was constructed by 
co-adding 19 good-quality images and all other images were subtracted from this 
master frame to produce the difference images. The next step was to carry out 
aperture photometry on the difference images using the centroids from the 
profile-fitting photometry of the reference image. The result was a time series 
of 4894 magnitude differences 
$\Delta m_{i} = -2.5\log [(F_0-\Delta F_i)/F_0)]$ for each star in the master 
list for the field, where $F_0$ and $\Delta F_i$ are the aperture fluxes of the 
star in the master and the $i$th difference image, respectively. The last step 
was to decorrelate the light curves by regressing each light curve against the  
light curves of $\sim 250$ other stars, a technique very similar to that 
described in \citet{Jenk00}. Average light curves were constructed by binning 
the decorrelated data in 0.0062~day ($\sim$9~minute)-wide bins. The typical 
scatter in the photometry of the brightest stars, as measured by the rms, is 
$\sim 0.005$~mag before and $\sim 0.002$~mag after the binning.

About 8000 stars with low enough scatter (RMS $\leq 0.02$~mag after the binning) 
were searched for transit events using the box least squares (BLS) algorithm 
\citep{Kova02}. The algorithm assumes only two levels in the periodic light 
curve, and thus is more efficient than generic methods that search for any 
periodic variation. It performs very well in low signal-to-noise ration (S/N) 
situations and is generally capable of detecting transit signals with amplitude 
of about 6 times the rms or better. The light curves of the eight transit 
candidates identified by the BLS search were examined visually to eliminate 
possible false detections such as stars with signs of a secondary eclipse or 
continuously varying light curve. We estimate that the visual inspection can 
detect secondary eclipses or modulated light curves with amplitudes as low as 
0.006--0.01~mag.

\begin{figure}
\caption{Binned $R$ light curve of \GSC\ obtained from the wide-field 
photometry. The data are folded with the photometric period of $3.35$ days. Only 
the portion near the transit times is shown. The rms over the whole observing 
season is $\sim 0.0028$~mag (see Section~\ref{sec:widefield}). \label{f:lc181}}
\plotone{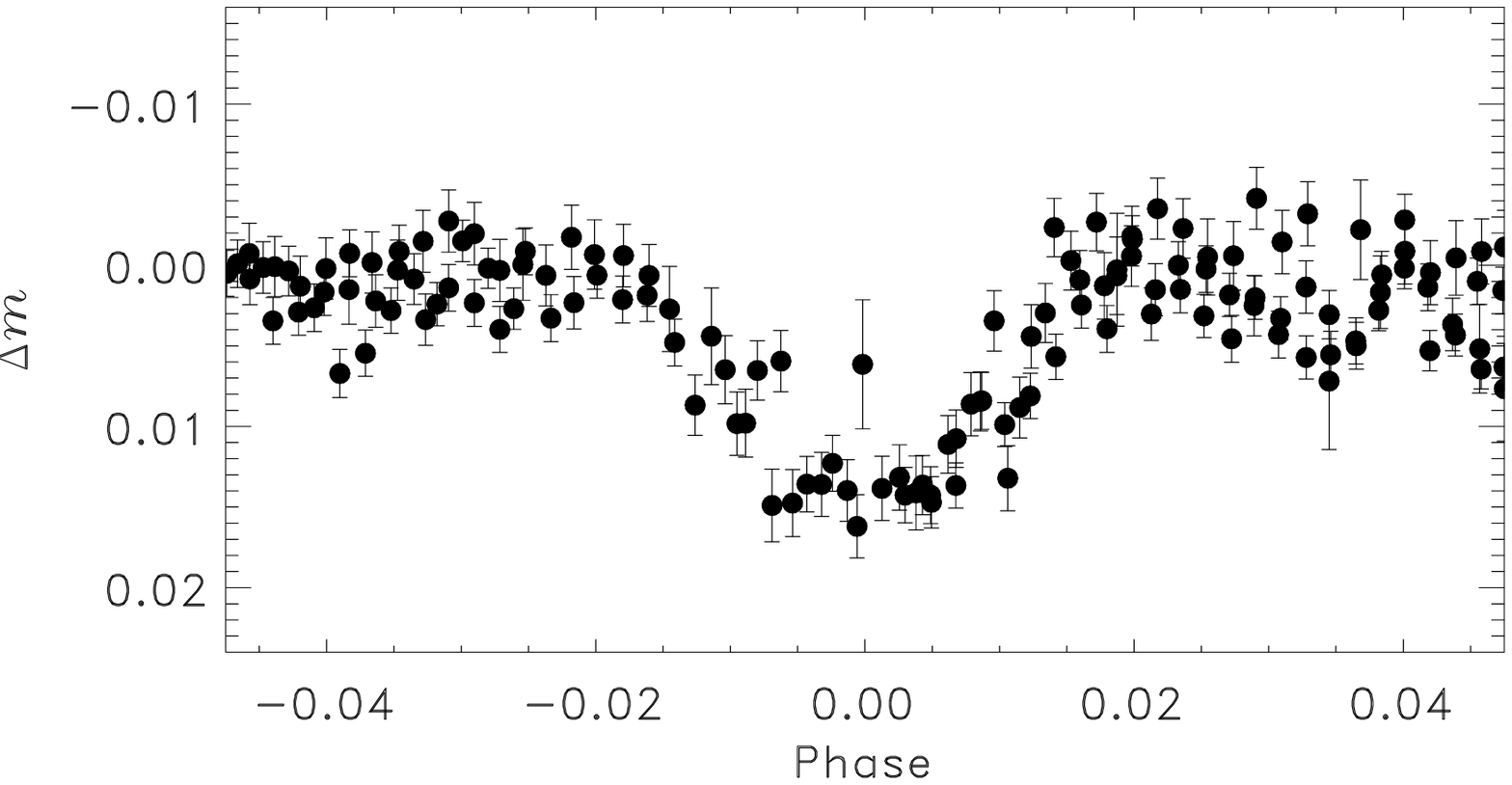}
\end{figure}

One of the best transit candidates was \GSC\ 
(J2000 coordinates: $\alpha = 08^{\rm h} 26^{\rm m} 22\fs 56$, 
$\delta = +26\arcdeg 59\arcmin 45\farcs5$). Its binned light curve, folded with 
the photometric period of $3.35$ days, is shown in Figure~\ref{f:lc181}. Each 
filled circle in the plot is the weighted average of up to five individual 
measurements. Two complete and two partial transit-like events with a depth of 
1.4\% and duration of 2.7 hours were observed. The Two Micron All Sky Survey 
(2MASS) colors of the object were typical for an F dwarf and examination of the 
Digitized Sky Survey\footnote{The Digitized Sky Survey was produced at the Space 
Telescope Science Institute under U.S. Government grant NAG W-2166. The images 
of these surveys are based on photographic data obtained using the Oschin 
Schmidt Telescope on Palomar Mountain and the UK Schmidt Telescope. The plates 
were processed into the present compressed digital form with the permission of 
these institutions.} images did not show any bright companions inside the 
$\sim 20\arcsec$ PSST stellar profile. Based on the shape and duration of the 
transit, and the preliminary spectral type from the 2MASS catalog, \GSC\ was 
classified as a probable transit candidate and was included in the 
radial-velocity monitoring program with the Harvard-Smithsonian Center for 
Astrophysics (CfA) Digital Speedometers \citep{Lath85,Lath92}. We also 
initiated a program of photometric and spectroscopic observations with larger 
telescopes, both to try to improve the photometric precision and to determine 
whether the transit-like events were truly caused by a dark substellar object.

\section{Multicolor Photometry \label{sec:multicolor}}

The principal motivation for obtaining multicolor photometry was two-fold: to 
check for color-dependent depth of the transit (which in some cases would 
indicate a stellar and not a substellar companion) and to obtain standard 
magnitudes and colors for the primary star. In addition we wanted to confirm the 
stellar parameters derived from the low-resolution CfA spectra that were already 
available at that time. 

We obtained off-transit $BV(RI)_{\rm C}$ exposures on 2004 March 26 and 
in-transit $BVI_{\rm C}$ exposures on 2004 March 27 (UT dates) with the Lowell 
Observatory $42\arcsec$ (105-cm) Hall reflector in combination with a 
$2{\rm K} \times 2{\rm K}$ SITe CCD. Aperture photometry was carried out on all 
images to derive the instrumental magnitudes. We used the mean magnitude of 
several comparison stars in each exposure to correct for the effects of 
atmospheric extinction, varying seeing, etc., and to produce the differential 
$BVI_{\rm C}$ light curves for \GSC. During the off-transit observations on 2004 
March 26, we also observed four photometric standards in the SA101 and G163 
fields \citep{Land92} in order to calibrate the photometry.

We also obtained in-transit $VI_{\rm J}$ photometry on 2004 April 2 with the 
Instituto de Astrof{\'\i}sica de Canarias (IAC) 82-cm telescope equipped with a 
$1{\rm K} \times 1{\rm K}$ CCD. Because of the smaller field of view 
($7\arcmin \times 7\arcmin$), only one comparison star was used to obtain the 
final differential photometry. The mean error of a single observation was 
$\sim 0.007$~mag in $V$\/ and $\sim 0.006$~mag in $I$.

\begin{figure}
\caption{Light curves of \GSC\ obtained from the multicolor photometry of 
the transits on 2004 March 27 and April 2. The data in consecutive filters are 
offset by $-0.05$ for clarity. The typical error of a single point is 
$\sim 0.004$~mag. \label{f:mcphot}}
\plotone{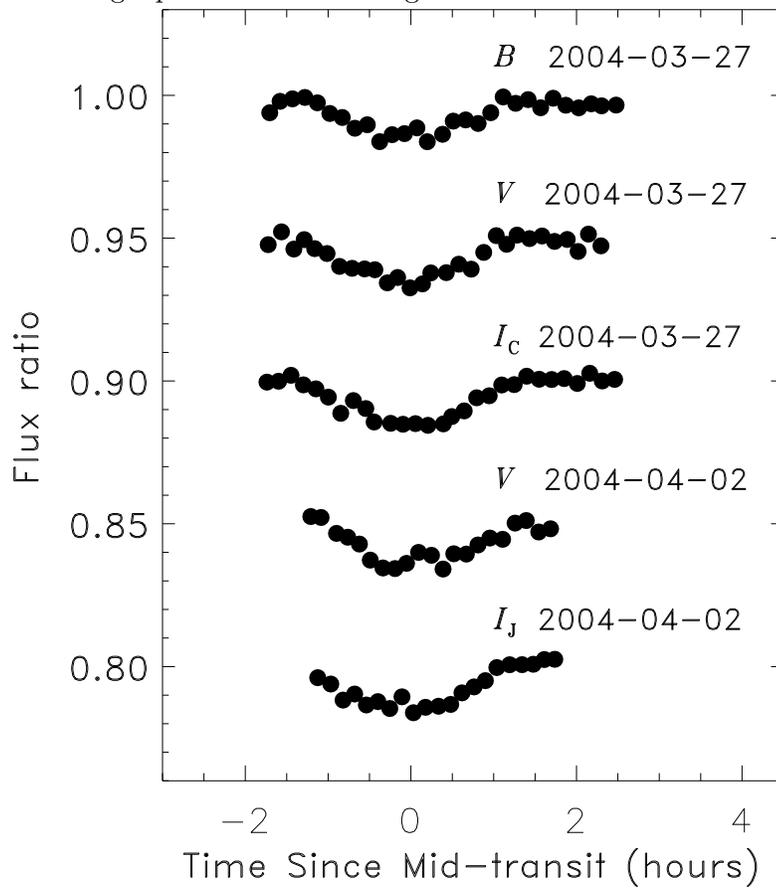}
\end{figure}

The transit light curves from 2004 March 27 (Lowell) and 2004 April 2 (IAC) are 
shown in Figure~\ref{f:mcphot}. Each plot is labeled with the filter and date of 
observation. The data were binned in the same way as the wide-field photometry, 
i.e., using bin widths of 0.0062 days; therefore each point in the plot is 
the weighted average of up to three individual measurements. On both nights the 
observations were affected by cirrus and the moon. Nevertheless, the photometry 
is good enough to ascertain that the depth of the transit-like events is equal 
within the observational errors in all three filters. A statistically 
significant difference in the transit depths (after accounting for limb 
darkening) would have rejected this object as a transit candidate. Equal depths,  
however, do not preclude the possibility of stellar companions of similar colors 
and low luminosity. The precision of our multicolor photometry is not enough to 
detect the effect of limb darkening with certainty: the expected transit depth 
difference for a solar-type star is $\sim 0.003$~mag between the $B$\/ and 
$I$\/ light curves \citep{Seag03} and could be even less for a hotter F star.

The observations of two full transits roughly a year after the PSST photometry 
from 2003 also allowed us to refine the system's photometric period and derive 
an improved transit ephemeris:
\begin{equation}
{\rm Min\ I\ (HJD)} = (2452676.3021 \pm 0.0005) + (3.35002 \pm 0.00002)\/E
\end{equation}
where $E$ is the number of whole periods elapsed since the reference transit 
epoch.

In Table~\ref{t:phot_data} we have summarized the principal properties of the 
transit candidate as derived from our wide-field and multi-color photometry, as 
well as the $JHK$ photometry from the 2MASS catalog. 

\section{Spectroscopic Observations \label{sec:spectroscopy}}

The transit candidate was observed spectroscopically at high resolution and high 
S/N at the Keck Observatory, and also with similar resolution but lower S/N at 
the CfA for the purpose of monitoring the radial velocity of the star. We 
describe these observations below.

\subsection{Keck Observations and Results \label{sec:keck}}

Observations of \GSC\ with the Keck~I telescope were obtained on 2004 April 3 
during the course of another program using the HIRES instrument \citep{Vogt94}. 
Two exposures of 10 and 20 minutes were obtained, together with exposures of a 
telluric standard and three spectral standards with accurately known radial 
velocities (see Table~\ref{t:hires_std}). The instrument was used with the red 
collimator and an RG-610 filter at cross-disperser angle of $1\fdg4005$ and 
echelle angle of $-0\fdg2860$. The D1 decker ($1\farcs15 \times 14\farcs0$) has 
a projected width of 4 pixels and produced spectra at 
$\lambda/\Delta\lambda = 34,\!000$. Sixteen echelle orders, each with a length 
of $\sim 100$~\AA, were recorded in the wavelength interval 6330--8750~\AA, with 
gaps between orders. Exposures of an internal quartz lamp for flat-fielding and 
a ThAr lamp for wavelength calibration were also made with the same setup. The 
spectra were extracted using the {\sc makee} package written by Tom Barlow, and 
then resampled onto a uniform logarithmic wavelength scale. The radial and 
projected rotational velocities of \GSC\ were determined via a cross-correlation 
analysis with the three spectral standards listed in Table~\ref{t:hires_std}, 
resulting in values of ${\rm RV} = +31.4 \pm 0.4~\kms$ and 
$v \sin i = 33.8 \pm 0.6~\kms$. For the cross correlations, we used 10 spectral 
regions of length 20--30~\AA\ that contained no prominent telluric absorption 
features or strong gravity sensitive features like K~{\sc I} at 7665 and 
7699~\AA, Na~{\sc I} at 8183 and 8195~\AA, or Ca~{\sc II} at 8498~\AA.

To estimate stellar parameters from the HIRES spectra, we obtained the atlas of 
synthetic spectra (based on the Kurucz ATLAS9 models) presented recently by 
\citet{Zwit04}. We used a restricted grid of parameters, spanning
$5000 < T_{\rm eff} < 7500$, $2.5 < \log g < 5.0$, and 
$-0.5 < [{\rm M}/{\rm H}] < 0.5$, where $T_{\rm eff}$ is in K and the surface 
gravity $\log g$ is in ${\rm cm\, s^{-2}}$. Microturbulent velocity was fixed at 
$2~\kms$ and the rotational velocity $v\sin i$ at $30~\kms$, the model 
grid-point closest to the estimated value of $34~\kms$. We repeated the analysis 
described below with models at $40~\kms$ and did not derive significantly 
different results. We interpolated the model spectra onto the wavelength 
solution for the HIRES data corrected for the system velocity. Telluric 
absorption features were identified in the observed spectrum of a 
rapidly-rotating A-star calibrator, and data at these wavelengths were given 
zero statistical weight in the subsequent analysis.

\begin{figure}
\caption{Two portions of the observed HIRES spectrum ({\em diamonds}) of \GSC\ 
in the vicinity of the Ca~{\sc II} infrared triplet. The solid line shows the 
best-fit model spectrum ($T_{\rm eff} = 6600$~K, $\log g = 4.3$, and 
$[{\rm M}/{\rm H}] = 0.0$). Two additional models are overplotted: 
$T_{\rm eff} = 6000$~K, $\log g = 4.3$, $[{\rm M}/{\rm H}] = 0.0$ ({\em dotted 
line}), and $T_{\rm eff} = 6600$~K, $\log g = 3.5$, $[{\rm M}/{\rm H}] = 0.0$ 
({\em dashed line}). These fits illustrate the sensitivity of features to 
$T_{\rm eff}$ and $\log g$, most noticeably in the wings of the Ca~{\sc II} 
feature at 866.2~nm.} \label{f:hires}
\plotone{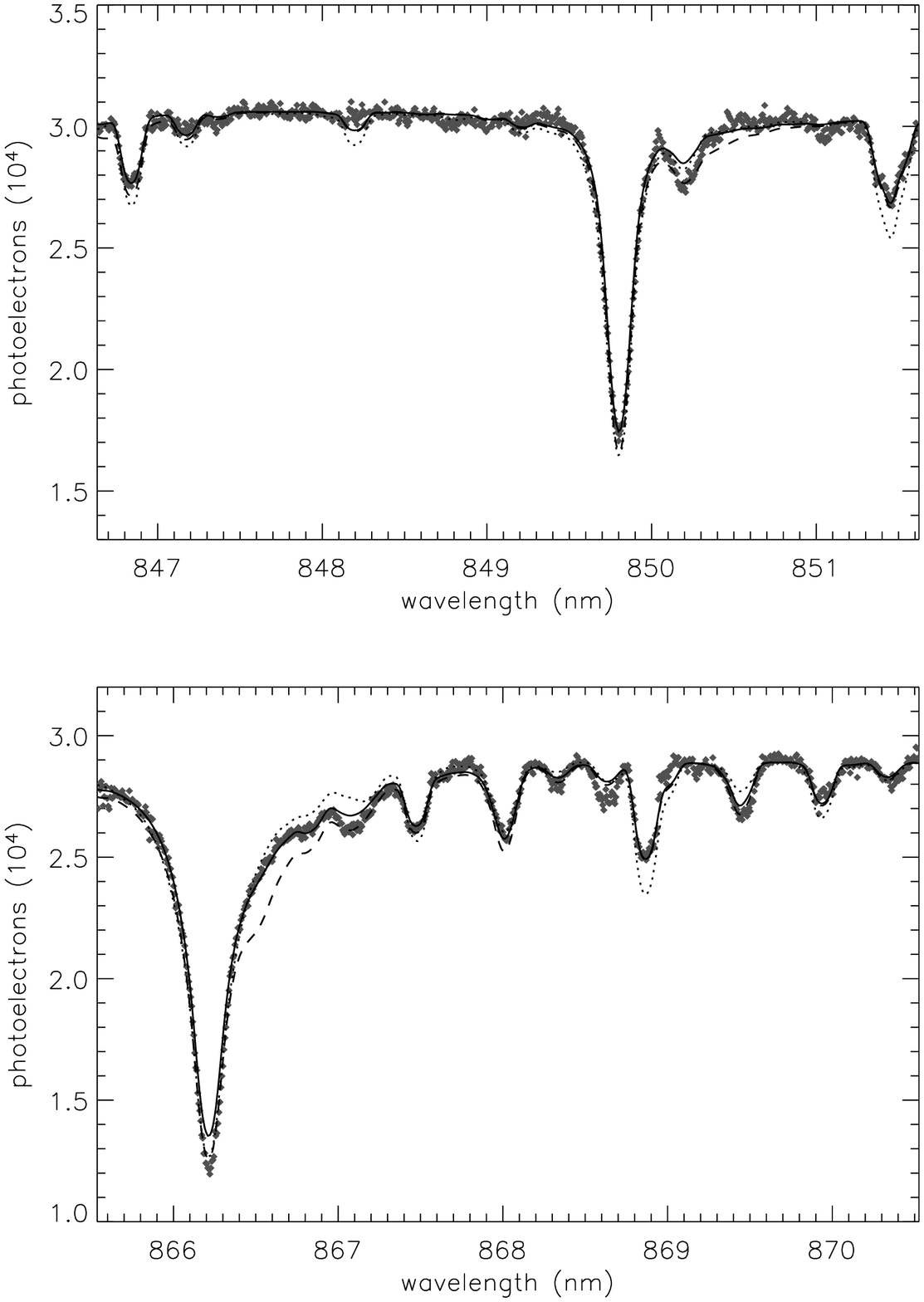}
\end{figure}

Five HIRES echelle orders overlapped with the spectral range of the models and 
were relatively free of telluric absorption: 7679--7791~\AA, 7850--7964~\AA, 
8028--8145~\AA, 8410--8533~\AA, and 8615--8741~\AA. For each of these 
echelle orders $n$, we calculated the $\chi^2$ difference 
$\chi^{2}_{n} \, (T_{\rm eff}, \log g)$ between each model spectrum (multiplied 
by a low-order polynomial to correct for the residuals in flattening the blaze 
function of the spectrograph) and the observed data. We then summed the 
$\chi^{2}_{n}$ values at each $T_{\rm eff}$ and $\log g$ over the five spectral 
orders $n$. We also considered models at $[{\rm M}/{\rm H}] = \pm 0.5$, both of 
which yielded significantly poorer fits to the data. The majority of the 
spectral features constrain primarily the temperature. However, our spectral 
range also encompasses two lines (at 8498 and 8662~\AA) from the Ca~{\sc II} 
triplet that yield good constraints on the surface gravity. Two portions of 
the spectrum in the vicinity of the Ca triplet features are shown in 
Figure~\ref{f:hires} and overplotted with model spectra.

The formal $\chi^2$ errors are not representative of the true uncertainties, 
since the theoretical spectra reveal systematic differences from the observed 
data (due to missing features or uncertainties in the line parameters). To 
evaluate the uncertainties in the stellar parameters, we repeated the above 
analysis for two of the spectral standard stars in Table~\ref{t:hires_std}, 
HD~78154A and HD~84737. We first rotationally broadened the observed spectra to 
30~\kms\ to simulate any degradation in the parameter estimation for \GSC\ due 
to its rapid rotation. Based on the accuracy of the model fits to these spectral 
standards, and a visual inspection of the model fits (see Figure~\ref{f:hires}), 
we estimate $T_{\rm eff} = 6600 \pm 150$~K and $\log g = 4.3 \pm 0.5$ for \GSC, 
consistent with an F5V star ($M \simeq 1.3M_\sun$, $R \simeq 1.3R_\sun$). 

\subsection{CfA Observations and Results \label{sec:cfa}}

Spectroscopic monitoring of \GSC\ was conducted at the CfA with the 1.5-m 
Tillinghast reflector at the F. L. Whipple Observatory (Mount Hopkins, Arizona), 
and the 1.5-m Wyeth reflector at the Oak Ridge Observatory (Harvard, 
Massachusetts), over a period of 3 months between 2004 February and May. 
Identical echelle spectrographs with photon-counting Reticon detectors were used 
on both telescopes, giving a resolving power of 
$\lambda/\Delta\lambda = 35,\!000$. A single echelle order spanning 45~\AA\ was 
recorded at a central wavelength of 5187~\AA, which includes the Mg~b triplet. A 
total of 33 spectra were obtained, with S/Ns ranging from about 14 to 35 per 
resolution element of 8.5~\kms. 

Radial velocities were measured by cross-correlation using the IRAF task 
{\sc xcsao} \citep{Kurt98}. The template was selected from a large library of 
synthetic spectra based on model atmospheres by R.\ L.\ 
Kurucz,\footnote{Available at http://cfaku5.cfa.harvard.edu.} computed for us 
by J.\ Morse. The parameters of the template are theeffective temperature, 
rotational velocity, and surface gravity, with the first two affecting the 
velocities the most. Solar metallicity was assumed throughout. The values best 
matching the star were determined from grids of cross-correlations, seeking the 
highest correlation averaged over all exposures. The formal values are 
$T_{\rm eff} = 6200 \pm 200$~K, $\log g = 3.5 \pm 0.5$, and 
$v\sin i = 34 \pm 2$~\kms, although temperature and gravity are strongly 
correlated owing to the narrow spectral window. The rotational velocity agrees 
well with the estimate from the Keck spectra, while the temperature is somewhat 
lower. In view of the correlation mentioned above we adopted a compromise value 
of $T_{\rm eff} = 6500$~K, corresponding roughly to an F5 star. For the gravity 
we adopted $\log g = 4.0$. The radial velocities we obtained, using the 
synthetic template from our library closest to these parameters, have internal 
errors around 1~\kms, which in our experience is quite typical for stars with 
this rotational broadening. Instrumental shifts affecting the velocities were 
monitored and corrected for by obtaining frequent exposures of the twilight sky, 
as described by \citet{Lath92}. 

\begin{figure}
\figcaption{Spectroscopic orbit adjusted to the velocities of \GSC\ using 
one-dimensional cross-correlations, i.e., under the assumption that the 
spectrum is single-lined. A circular orbit is assumed, and the ephemeris (period 
and epoch) are adopted from Eq.[1]. The dotted line represents the 
center-of-mass velocity. The velocity semi-amplitude of 3.55~\kms\ implies a 
brown dwarf companion with a mass of 32~$M_{\rm Jup}$ (see text), assuming a 
primary mass of $1.3~M_\sun$. \label{f:orbit1}}
\plotone{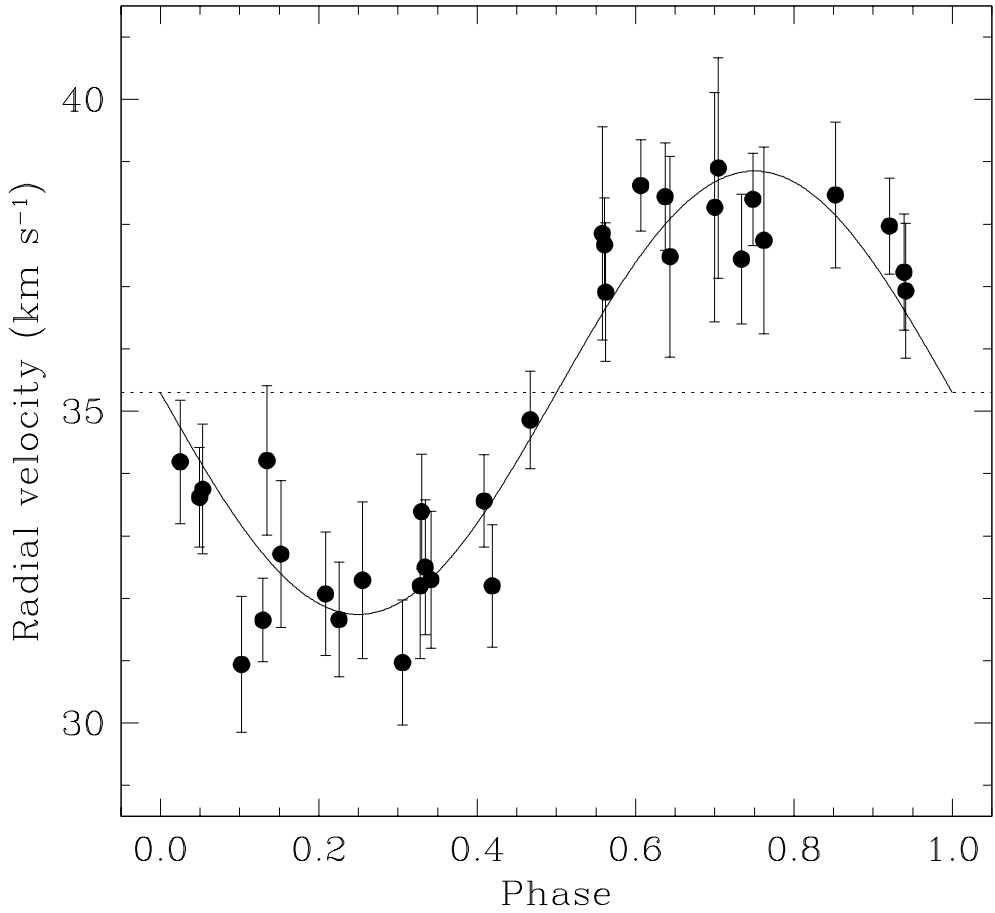}
\end{figure}

The radial velocities show a clear pattern of variation with the same period as 
the photometry (Eq.[1]), and a phasing that is consistent with what is expected 
for transits (negative slope at phase 0.0, which corresponds to the epoch of the 
transit). This suggests orbital motion. A Keplerian orbit (assumed to be 
circular owing to the short period) adjusted to these data is shown in 
Figure~\ref{f:orbit1}, where the ephemeris has been adopted from photometry and 
held fixed. The velocity semi-amplitude of $K = 3.55 \pm 0.21$~\kms\ is highly 
significant. If due to orbital motion, the minimum mass of the companion is 
determined to be $M_2 \sin i = (0.0250 \pm 0.0030) (M_1+M_2)^{2/3}$~$M_\sun$, 
where $M_1$ is the mass of the star and $i$ is the inclination angle of the 
orbit. With a typical mass for an F5 main sequence star of 
$M_1 \approx 1.3M_{\sun}$ and an orbit that is essentially edge-on as implied 
by the transits, this corresponds to a companion mass of 
$\sim 32$~$M_{\rm Jup}$, which is clearly in the brown dwarf regime. The 
center-of-mass velocity of the system is $+35.30 \pm 0.15$~\kms, and the rms 
residual from the fit is 0.88~\kms. 

While the discovery of a bona-fide brown dwarf companion to a solar-type star in 
such a tight orbit would certainly be of great interest, finding one that also 
transits across the disk of its parent star would be unprecedented and would 
allow for the first time the measurement of the size and mean density of such an 
object directly. The observational evidence presented above for \GSC\ is fairly 
typical of what other transit surveys and subsequent follow-up work might 
produce, and no obvious signs are seen that would indicate a false positive 
detection (see also Section~\ref{sec:recovering}). For example, visual 
examination of the spectra and of the corresponding cross-correlation functions 
revealed no sign of lines from another star that might be causing spurious 
velocity variations. Furthermore, there is no significant wavelength dependence 
to the depth of the transits (see Section~\ref{sec:multicolor}), which would be 
expected in general if more than one star is involved. In view of the importance 
of this case, however, we undertook a series of other tests described below. 
These ultimately proved that the star is, in fact, \emph{not} orbited by a brown 
dwarf companion, but instead that both the photometric and the spectroscopic 
signatures are the result of blending with an eclipsing binary. We present these 
tests in some detail as an example of the care required to validate transit 
candidates that are being generated by these kinds of searches. 
	
\section{Identifying A False Positive \label{sec:false}}

One of the potentially observable manifestations of a blend scenario is periodic 
changes in the shapes of the spectral lines due to contamination from the 
eclipsing binary, which may give the impression of radial velocity variations in 
the brighter star. Although different in nature, this is qualitatively similar 
to the effect produced by chromospheric activity in more active stars, which can 
also produce spurious velocity variations \citep[see, e.g.,][]{Quel01}. 

Direct spectroscopic detection of the eclipsing binary in \GSC\ may be more 
difficult, particularly if it is faint. However, additional information on the 
properties of the configuration can be obtained from detailed modeling of the 
light curve, and this can significantly constrain the problem and aid in the 
disentangling of the spectral features of the binary, even allowing the 
measurement of its radial velocity. We focus below on the CfA spectroscopy that 
provides the better time coverage, and we explore these issues in turn. 

\subsection{Line Shapes}

\begin{figure}
\caption{({\em a}) Line bisectors for each of our CfA spectra of \GSC. The 
filled circles on each bisector at correlation values of 0.0 and 0.75 indicate 
the reference levels adopted here for computing the span in velocity. ({\em b}) 
Bisector span used as a proxy for line asymmetry, as a function of orbital phase 
(see text). One spectrum shows significant contamination by moonlight, and has 
been omitted. The obvious variations strongly suggest blending with another 
star. \label{f:bisectors}}
\plotone{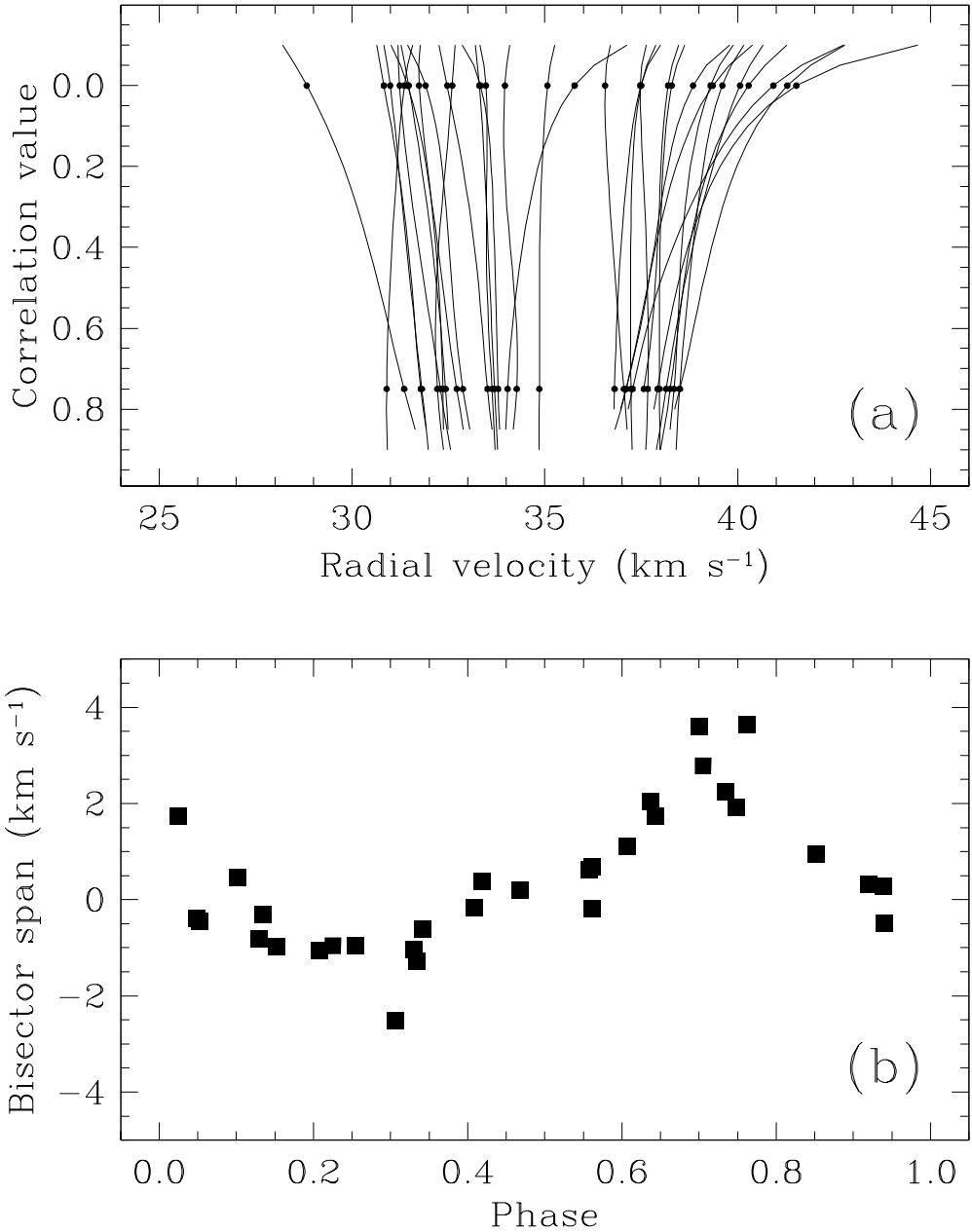}
\end{figure}

Beyond the visual inspection of the spectra and cross-correlation functions for 
\GSC\ described in the previous section, a much more sensitive measure of the 
presence of contaminating lines from another star can be obtained by examining 
the profiles of the spectral lines for asymmetries. These asymmetries can be 
quantified by measuring the ``bisector spans," usually defined as the velocity 
difference between the top and bottom of the line bisectors 
\citep[see, e.g.,][]{Quel01,Torr04a} or between any other two reference levels. 
We have computed these here for our CfA spectra directly from the 
cross-correlation functions, which can be taken to represent the profile of the 
average spectral line in the star. Figure~\ref{f:bisectors} shows the 
individual line bisectors from our spectra, along with the bisector spans 
plotted as a function of the photometric phase, based on the ephemeris from 
Eq.[1]. We note that the magnitude of the asymmetries in \GSC\ is substantially 
larger than in the Sun, where lines typically show spans of only a few hundred 
meters per second. The rather obvious trend with phase is a strong indication of 
contamination by the lines of another star, moving back and forth with respect 
to the (presumably) fixed lines of the F star with a period of 3.35 days. Thus, 
changes in the line asymmetry appear to have resulted in spurious velocity 
variations due to the measuring technique, which uses a significant portion of 
the line profile.\footnote{The radial velocity is determined in {\sc xcsao} as 
the centroid of the peak of the cross-correlation function, from a quadratic fit 
to the top 50\% of the peak. Thus, asymmetries in the profile will affect the 
centroid directly.} The pattern of the variations (first negative, then 
positive) is the same as that of the velocities in Figure~\ref{f:orbit1}. The 
implication is that the line-profile changes are due to the primary of an 
eclipsing binary along the same line of sight. Blending with the brighter star 
dilutes the normal eclipses and reduces their depth to the level that we see, 
only 1.4\%. 

\subsection{Modeling The Light Curve As A Triple Star System}
\label{sec:modeling}

To constrain further the properties of the blend and derive properties for the 
eclipsing binary that will help in the direct detection of its lines in the 
spectra of \GSC, we turn next to the photometry. We model the $R$-band light 
curve as a combination of the light from an eclipsing binary plus a brighter 
third star, following \cite{Torr04b}. Briefly, the light variations (not only 
the depth and duration of the transit, but also the detailed shape) are modeled 
with a combination of three stars whose properties are parameterized in terms of 
their mass. All other stellar properties are derived from model isochrones, 
which can be the same for the eclipsing binary and the third star if they form a 
physical triple system, or different otherwise. The mass of the brightest star 
in the system is constrained by the spectroscopic and photometric information 
available (see Section~\ref{sec:spectroscopy}), mainly the effective temperature 
of 6500~K. These light-curve fits are usually not unique. However, once the 
masses for the stars composing the eclipsing binary are determined from the fit, 
other properties of the blend can be predicted that can potentially be tested 
against the observations, such as the brightness of the eclipsing binary 
relative to the brightest star (amount of dilution), the orbital velocity 
amplitudes of the eclipsing binary, and the projected rotational velocities of 
its components \citep[for details see][]{Torr04b}. As we show below the powerful 
combination of the photometric and spectroscopic constraints typically resolves 
the ambiguities, and allows one to arrive at a single configuration that agrees 
with all the observations. 

Light-curve modeling of a physical triple system (all stars at the same 
distance) assuming that the brightest star is an F5 main-sequence object 
resulted in a good fit to the observations but implied an eclipsing binary that 
is far too bright to be missed in our spectra. Placing the binary in the 
background to make it fainter (thus removing the condition of physical 
association with the F star) resulted in fits that were also acceptable. 
However, the near-sinusoidal pattern of line asymmetries in 
Figure~\ref{f:bisectors} suggests that if there is another star in the spectrum, 
its center-of-mass velocity must be fairly close to the mean velocity of \GSC. 
Although not impossible, this would be rather unlikely for a line-of-sight 
alignment of unrelated stars. We, therefore, explored other possibilities in 
which the F star is intrinsically brighter. 

The spectroscopic constraint we have on the surface gravity of \GSC\ is such 
that a modest degree of evolution cannot be ruled out by the observations (see 
Section~\ref{sec:spectroscopy}). Because of the morphology of the isochrones in 
the vicinity of the turnoff, it is possible within certain ranges of age and 
temperature for an evolved F star to be significantly brighter at any given 
temperature (by up to 2 mag in some cases) if it is near the end of its 
main-sequence life, beyond the turnoff point. The mass will obviously be 
considerably larger than a star located before the turnoff at the same 
$T_{\rm eff}$ (see Figure~\ref{f:iso} below). Our estimate of the effective 
temperature for \GSC\ happens to be in the range that allows this ambiguity. 

\begin{figure}
\figcaption{Light-curve fit to the wide-field $R$-band photometry of \GSC, under 
a blend scenario with an eclipsing binary (G0 star + M3 star) diluted by the 
light of an F5 star (see text). The bottom panel shows an enlargement in the 
vicinity of the primary eclipse. A shallow secondary eclipse is predicted 
({\em top panel}) but is undetectable in the present data. \label{f:blendfit}}
\plotone{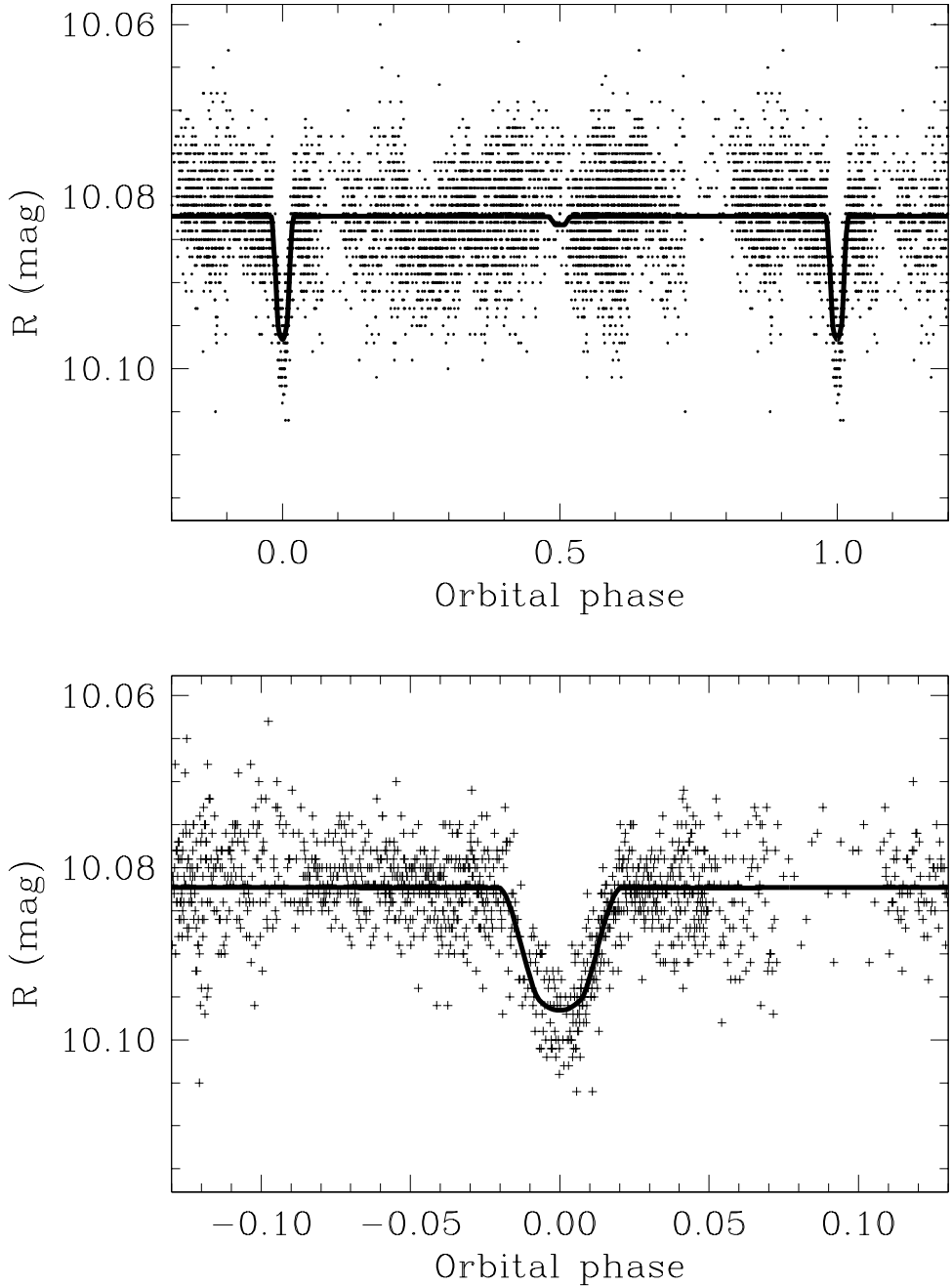}
\end{figure}

Extensive light-curve modeling under this scenario resulted in a solution for a 
triple system providing an excellent fit to the photometry, with an eclipsing 
binary (composed of a G0 star orbited by an M3 star) roughly an order of 
magnitude fainter than the slightly evolved F5 star, consistent with the 
detection limits from our visual inspection of the spectra. This fit is shown in 
Figure~\ref{f:blendfit}. The secondary eclipse seen in the plot has a predicted 
depth of only 0.002 mag, which is undetectable in our present data. The 
inclination angle that provides the best solution is 87\arcdeg, and the F star 
contributes 89\% of the total light in the $R$ band. The properties of the three 
stars composing the blend are listed in Table~\ref{t:blendfit}, where the mass 
of the F star was fixed to give a temperature in close agreement with our 
spectroscopic estimate of 6500~K. The age determined for the system is 
$\sim$1.6~Gyr, and the distance (ignoring extinction and forcing agreement with 
the total apparent magnitude in the $R$ band) is approximately 500~pc. The 
location of the three stars in the H-R diagram is illustrated in 
Figure~\ref{f:iso}, along with model isochrones from the series by 
\cite{Gira00}. 

\subsection{Spectroscopically Recovering The Eclipsing System}
\label{sec:recovering}

Given the short orbital period, the stars in the eclipsing binary are likely to 
have their rotation synchronized with the orbital motion due to tidal forces. 
For solar-type stars this would result in a line broadening of some 15~\kms. In 
view of the rapid rotation of the F star itself ($v \sin i = 34$~\kms), we 
anticipate that the lines of the G star in the eclipsing binary will essentially 
be always blended with those of the F star. This will make their direct 
detection challenging (particularly if they are faint), even knowing where to 
look. As an illustration we show in Figure~\ref{f:twoccfs} the cross-correlation 
functions for two of our observations of \GSC\ near the quadratures, where the 
velocity separation between the G star and the F star should be the largest if 
the modeling in the preceding section is correct. We indicate with arrows the 
expected location of the G star (see below), but no obvious signs of it are 
seen. 

Modern techniques for analyzing composite spectra such as the two-dimensional 
cross-correlation algorithm TODCOR \citep{Zuck94} have the potential to enable 
this detection, even under severe line blending, thereby providing definitive 
proof of a blend. In TODCOR the observed spectra are cross-correlated against a 
composite template made by adding together two separate templates chosen to 
match each star. A two-dimensional correlation function is produced by exploring 
all possible combinations of Doppler shifts of the two stars. The maximum of 
this function should occur when the Doppler shifts of the two templates match 
the observation. This technique effectively decouples the velocities of the two 
stars, overcoming the blending problems of standard one-dimensional correlation 
techniques. Additionally, the light ratio can be left as an additional free 
parameter to be determined from the same observations \citep[see][]{Zuck94}. 

The modeling in Section~\ref{sec:modeling} predicts that the brightness of the 
G star (the primary star in the eclipsing binary) relative to the F star in 
\GSC\ is 0.12 in the $V$ band, which is close to the central wavelength of the 
CfA spectra. This corresponds to $\Delta V = 2.3$~mag. Additionally, its 
velocity semi-amplitude and rotational velocity are predicted to be 39.4 and 
16.8~\kms, respectively. These values, along with the temperature and 
surface gravity from Table~\ref{t:blendfit}, provide the information needed to 
look for direct signs of the G star in the spectra. Using a synthetic template 
closely matching these parameters, along with the same template adopted earlier 
for the F star, we re-analyzed the CfA spectra using TODCOR. We were indeed 
able to detect the G star, and measure its radial velocity (as well as that of 
the F star) in each of our spectra. These radial velocities are listed in 
Table~\ref{t:todcor}. An illustration of this detection is given in 
Figure~\ref{f:todcor}, in which the top panel shows a contour plot of the 
two-dimensional correlation function for one of our spectra at the second 
quadrature (same spectrum as in the bottom panel of Figure~\ref{f:twoccfs}). 
Cross sections through the maximum of the correlation function are indicated by 
dotted lines, and are displayed in the lower two panels. The peak at the 
expected location for the G star in the bottom panel shows that the lines of 
that star are present in the spectrum, and allow the Doppler shift to be 
measured. 

A Keplerian orbit adjusted to the velocities of the G star, with the ephemeris 
from Eq.[1] held fixed and assuming a circular orbit, gives a semi-amplitude of 
$38.24 \pm 0.80$~\kms. This is very close to the predicted value. The 
center-of-mass velocity of this orbit is $+35.71 \pm 0.59$~\kms, and the rms 
residual of the fit is 3.36~\kms. The residuals of the individual velocity 
measurements are listed in Table~\ref{t:todcor}, along with the orbital phase. 
The observations and fitted orbit are shown graphically in Figure~\ref{f:orbit2}.
Following \cite{Zuck94}, we also determined the light ratio directly from our 
spectra. The result, $0.12 \pm 0.02$, is virtually identical to the predicted 
value. 

The velocities of the F star measured with TODCOR, which previously displayed a 
clear variation with a semi-amplitude of 3.5~\kms\ (Section~\ref{sec:cfa}), are 
now essentially constant: the rms residual around the mean is 0.76~\kms\ (see 
Figure~\ref{f:orbit2}). This indicates that TODCOR has effectively removed the 
contamination from the primary of the eclipsing binary, which was biasing our 
original measurements. For reference, we include in Figure~\ref{f:orbit2} the 
spurious orbit from Figure~\ref{f:orbit1} ({\em dashed line}), which is seen not 
to fit the new measurements. We note also that the mean velocity of the F star 
as determined from its new velocities, 
$\langle{\rm RV}_{\rm F}\rangle = +35.40 \pm 0.13$~\kms, is the same as the 
center-of-mass velocity of the eclipsing binary, within the errors. For a 
hierarchical system this similarity may be indicative of a wide orbit for the 
eclipsing pair around the F star, unless that orbit has a low inclination angle 
or we happened to measure the system at a special phase. The lack of any 
significant variation in the velocities of the F star over the three-month 
interval of our observations implies an outer period that is probably 
significantly longer than this. 

With the eclipsing binary now confidently identified in our spectra, as a check 
we repeated the determination of the template parameters in the same way as done 
originally, but extending the technique to the case of composite spectra 
\citep[see, e.g.,][]{Torr02}. Although the results are more uncertain because 
twice as much information is being extracted from the same spectra (the 
effective S/N of each star is lower), the parameters we obtained for the F star 
($T_{\rm eff} = 6400$~K, $v \sin i = 33$~\kms) are very close to what we had 
been using, and those for the primary of the eclipsing binary 
($T_{\rm eff} = 5900$~K, $v \sin i = 18$~\kms) are also in excellent agreement 
with the blend model. We point out that the latter values (which are purely 
spectroscopic) are essentially independent of the light-curve modeling and thus 
provide gratifying confirmation of the blend results. 

\subsection{System Colors}

Our modeling of the $R$-band photometry allows us to predict the depth that the 
eclipses would have in other passbands, in particular in $B$, $V$, and 
$I_{\rm C}$, which we observed (see Section~\ref{sec:multicolor}). With the 
blend parameters described above, all three of these filters show the same 
expected transit depth as in the $R$ band to within about a milli-magnitude, a 
difference that is well below our detection threshold. The reason for the 
similarity is the small difference in color (or effective temperature) between 
the F star and the G star. Thus, \GSC\ represents a type of blend that is 
virtually impossible to expose from the wavelength dependence of the transit 
depth. 

Finally, an additional check for self-consistency in our model is given by the 
comparison between the predicted total apparent magnitudes for the blend (sum of 
three stars) and the measured brightness of \GSC\ in a variety of passbands 
spanning the optical and near-infrared. This comparison is shown in 
Table~\ref{t:mags}. The differences between the observed and predicted 
magnitudes are no larger than 0.02 mag, which is quite a remarkable agreement 
considering all the uncertainties involved, including systematic errors in the 
fluxes from the evolutionary models and other errors of an observational nature. 

\section{Discussion}

Current wide-field transit surveys easily achieve the required photometric 
precision and time coverage necessary to detect short-period transiting giant 
planets \citep[see, e.g.,][]{Bako04,Dunh04}. Our experience is that a single 
$6\degr \times 6\degr$ field yields anywhere from five to 20 or more transit 
candidates depending on the number of stars with sufficient photometric 
precision (from $\sim 3000$ in fields at high Galactic latitude to more than 
20000 in fields near the Galactic plane). The principal difficulty facing such 
surveys is the effective rejection of false alarms, such as the just described 
case of \GSC\ and others \citep[e.g.,][]{Char04}. Some false positives are 
relatively easy to uncover using high-cadence photometry or low-precision radial 
velocity measurements \citep{Lath03,Char04}. However, the case of \GSC\ shows 
that a blend of an eclipsing binary with a third star can mimic a transiting 
planet almost perfectly.

It has often been argued that blend configurations --- perhaps the most 
insidious of the false positive scenarios affecting transit candidates --- can 
usually be detected from the wavelength dependence of the transit depth. The 
case of \GSC\ is a poignant counterexample. As described above, the close 
similarity in color (temperature) between the two brightest components leads to 
little or no wavelength dependence. This underlines the importance of careful 
spectroscopic follow-up, not only to measure radial velocities but also to 
examine the spectral line profiles. Furthermore, it would appear that such cases 
may not be as rare as previously thought among certain kinds of stars. The 
transit candidate OGLE-TR-33, found in the course of the OGLE survey 
\citep{Udal02a}, was also shown by \cite{Torr04b} to be a blend resulting from 
a hierarchical triple system involving an eclipsing binary. In both \GSC\ and 
OGLE-TR-33, the brightest star is of spectral type F. This is significant, since 
\cite{Torr04b} showed (see their Fig.~10) that precisely in the F stars the 
main sequence widens enough that it allows an evolved star to be intrinsically 
brighter by up to 2~mag at a given temperature, leaving ample room to hide a 
fainter star (the eclipsing binary) in its glare. Additionally, F stars are 
typically rotating quite rapidly ($v \sin i \simeq 30$--40~\kms\ or more) and 
their line broadening is often similar to the orbital velocity amplitude 
expected for the eclipsing binary in a blend (for short periods typical of 
transit candidates). Under these conditions the spectral lines of the bright 
star and the eclipsing binary will be severely blended over much of the orbit, 
making it difficult to detect the latter. As a result the line asymmetries 
caused by contamination from the eclipsing binary will lead to spurious 
velocity variations measured for the bright star that are nearly sinusoidal, and 
therefore more easily confused with real variations. Thus, transit candidates 
that are F stars require extra care. Such candidates could be quite common, 
given that at the typical Galactic latitude 
($\left | b \right | \simeq 15\degr$) of our magnitude-limited surveys the 
maximum of the star counts of main-sequence stars is around spectral type F5. 

To date, five transiting extrasolar planets have been discovered from their 
photometric signatures and confirmed spectroscopically: one in a wide-field 
survey \citep[TrES-1,][]{Alon04}, and four around stars in the OGLE-III survey 
\citep{Udal02b,Udal02c,Udal03}: OGLE-TR-56 \citep{Kona03}, OGLE-TR-111 
\citep{Pont04}, OGLE-TR-113 \citep{Bouc04,Kona04}, and OGLE-TR-132 
\citep{Bouc04}. They have all undergone an analysis of the bisector variations 
and possible blend scenarios, and so the detection of planetary companions 
around those stars can be considered secure. Recent spectroscopic follow-up of 
several OGLE transit candidates \citep{Kona03,Torr04b} demonstrates that even 
with stars as faint as those ($V \sim 16$) the analysis technique applied here 
is able to distinguish between blends and true transiting planets. 

\acknowledgments

We thank Mike Calkins, Perry Berlind, and Joe Zajac for obtaining many of the 
CfA spectra. We also thank Tomaz Zwitter for allowing us access to his 
unpublished atlas of Kurucz stellar atmosphere models, Peter Stetson for 
providing us with the {\sc daophot~ii} package, and Phil Massey for the 
assistance with obtaining observing time on the $42\arcsec$ Hall telescope. 

Some of the observations presented here were obtained at the W. M. Keck 
Observatory, which is operated as a scientific partnership of the California 
Institute of Technology, the University of California, and the National 
Aeronautics and Space Administration (NASA). The observatory was made possible 
by the generous financial support of the W. M. Keck foundation. This work was 
carried out in part with support from grants NAG5-8271, NAG5-12088, and 
NNG04LG89G under the auspices of the NASA Origins of Solar Systems Program, 
and the Keck PI Data Analysis Fund (JPL 1260769).

This publication makes use of data products from the Two Micron All Sky Survey,
which is a joint project of the University of Massachusetts and the Infrared 
Processing and Analysis Center/California Institute of Technology, funded by 
the National Aeronautics and Space Administration and the National Science 
Foundation. This research has also made use of the Simbad database operated at 
Centre de donn\'ees astronomiques de Strasbourg, in Strasbourg, France.

\clearpage

\begin{figure}
\figcaption{Model isochrones from \cite{Gira00} for solar metallicity, for ages 
from 1 to 2.5~Gyr (in logarithmic steps of 0.05 in $\log t$, with $t$ in 
yr). The isochrone represented by the thick line was found to give the best 
fit in the blend modeling of \GSC\ and corresponds to an age of about 1.6~Gyr 
(see text). The dotted vertical line indicates the spectroscopic constraint on 
the effective temperature of the brightest star ($\sim$6500~K). We represent 
that star by the large filled circle near the end of the main sequence. The 
stars composing the eclipsing binary are represented by the two smaller filled 
circles on the same isochrone. \label{f:iso}}
\plotone{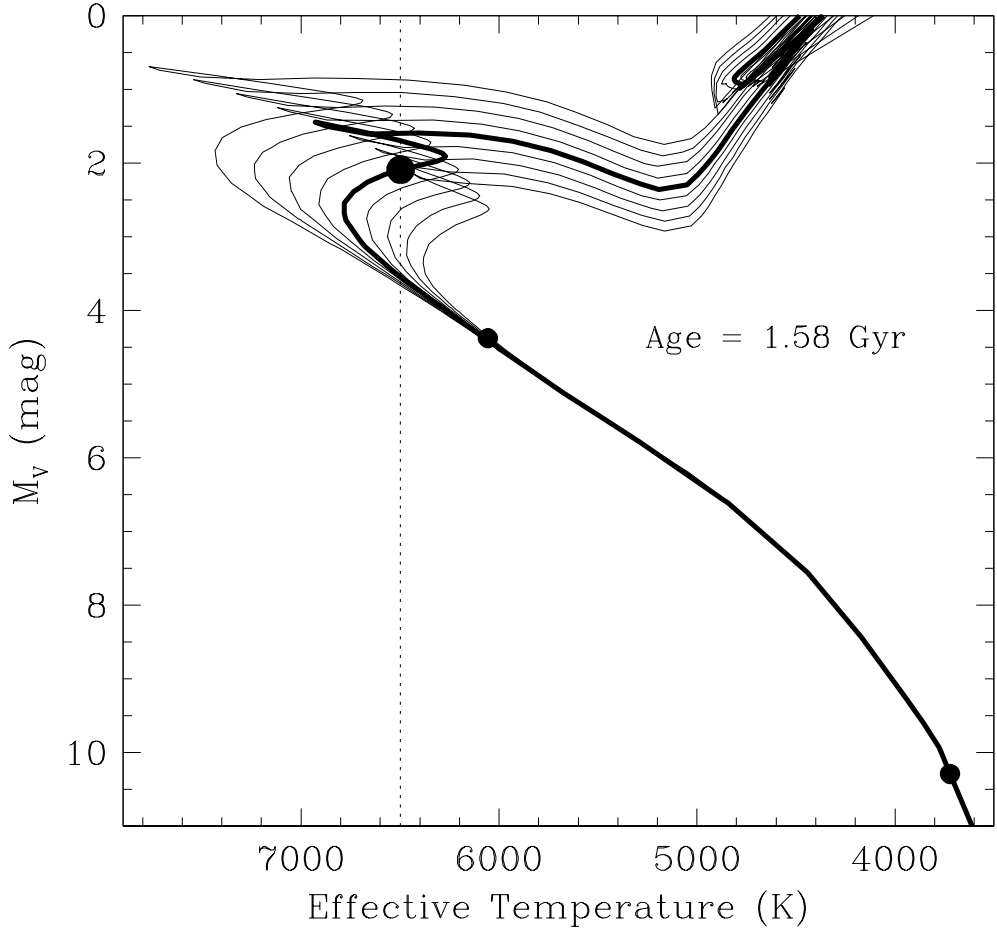}
\end{figure}

\begin{figure}
\figcaption{Cross-correlation functions from two of our spectra of \GSC\ 
obtained near the quadratures. An arrow indicates the expected location of the 
G star. No obvious sign of that star is seen in the wings of the broadened F 
star. \label{f:twoccfs}}
\plotone{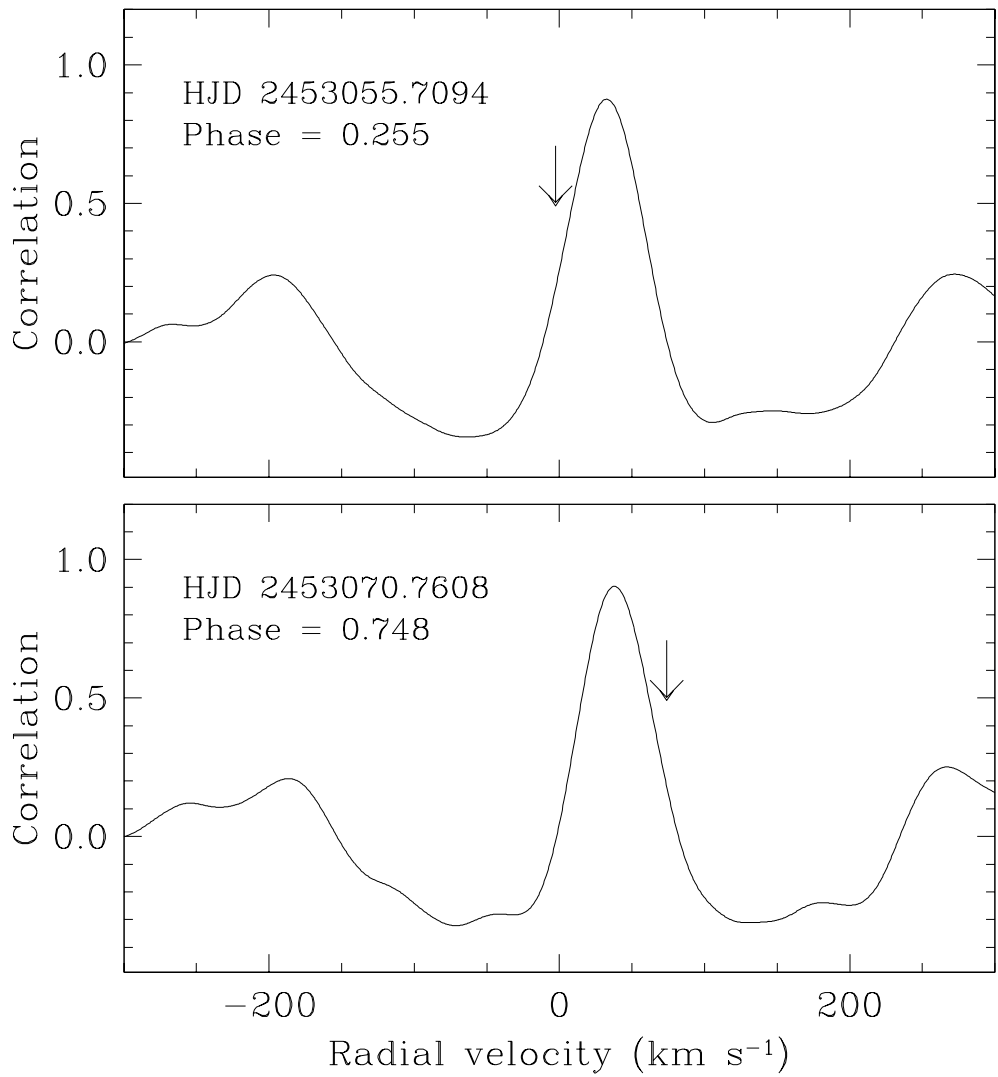}
\end{figure}

\begin{figure}
\figcaption{({\em a}) Contour diagram of the two-dimensional cross-correlation 
surface computed with TODCOR as a function of the velocities of the F star and 
the G star, for one of our spectra of \GSC\ near phase 0.75 (same spectrum as in 
the bottom panel of Figure~\ref{f:twoccfs}). The maximum is indicated by the 
filled circle, and the dotted lines represent cuts through the surface that are 
shown in ({\em b}) and ({\em c}). ({\em b}) Cross section of the two-dimensional 
correlation function at a fixed velocity for the G star, showing a peak 
corresponding to the F star in \GSC. Its measured velocity is represented by the 
dotted line.  ({\em c})  Same as ({\em b}), but at a fixed velocity for the F 
star. The peak corresponding to the G star (albeit noisy) is a clear sign of its 
detection. \label{f:todcor}}
\plotone{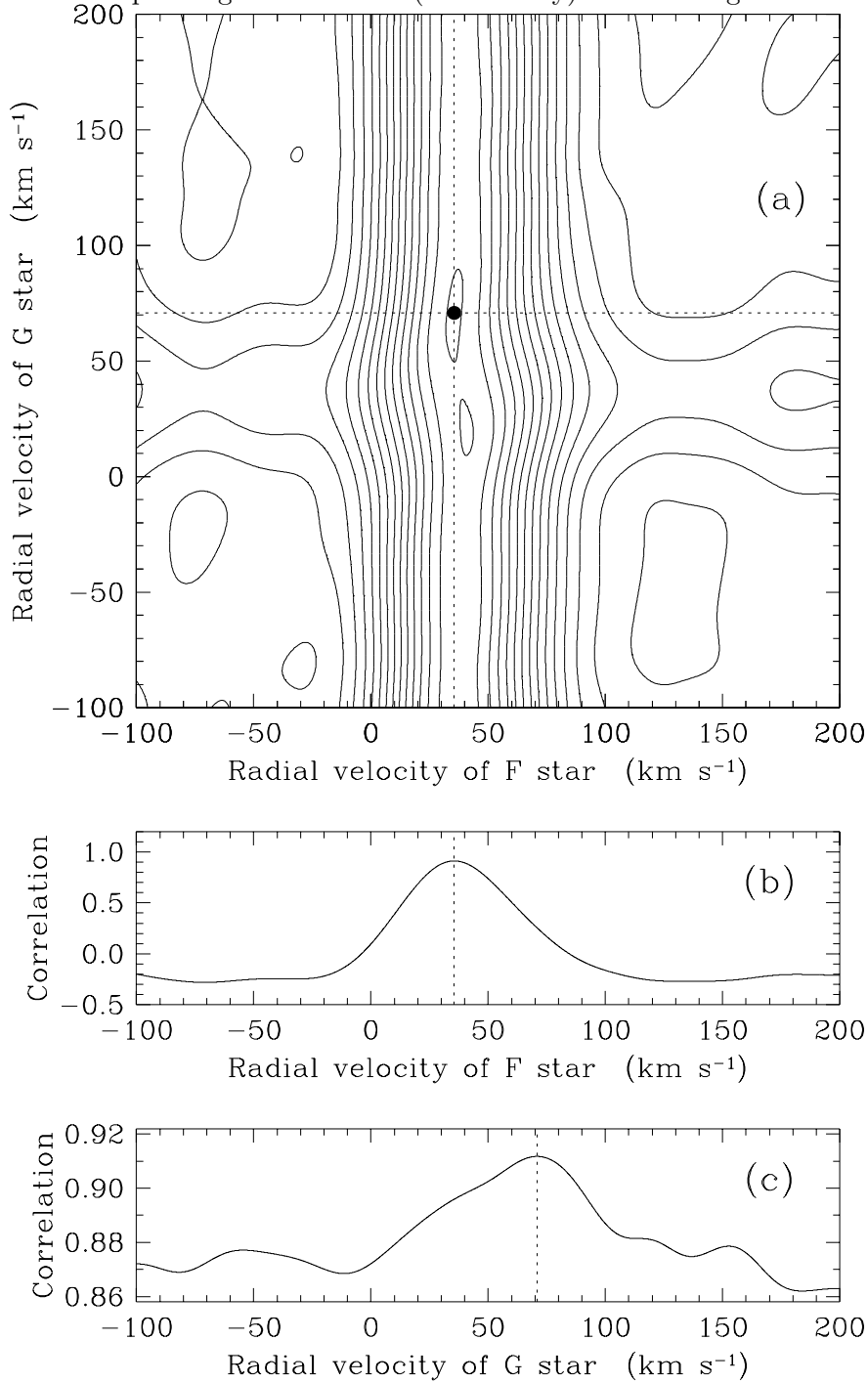}
\end{figure}
	
\begin{figure}
\figcaption{Radial velocity measurements from TODCOR for the G star in the 
eclipsing binary ({\em open circles}) and the F star ({\em filled circles}). A 
spectroscopic orbit has been fitted to the eclipsing binary, assuming zero 
eccentricity and adopting the ephemeris from Eq.[1]. Shown also for reference is 
the spurious orbit from Figure~\ref{f:orbit1} for the F star ({\em dashed line}), 
which is seen not to fit the new measurements. The center-of-mass velocity of 
the eclipsing binary is indicated by the dotted line and agrees with the 
velocities of the F star within the errors (see text). \label{f:orbit2}}
\plotone{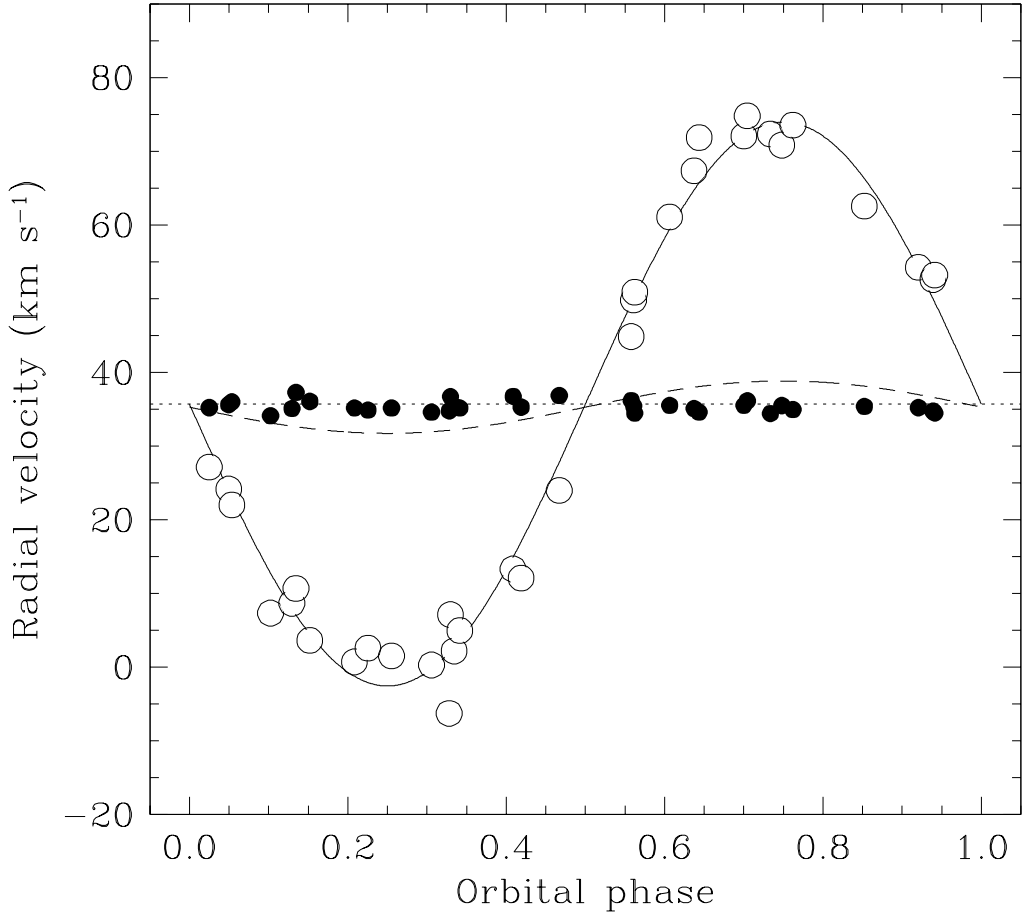}
\end{figure}

\clearpage

\begin{deluxetable}{lr}
\tablewidth{0pt}
\tablecaption{Data on \GSC\ \label{t:phot_data}}
\tablehead{
\colhead{General} & \colhead{Photometry}}
\startdata
GSC 01944-02289                                   & $V = \: 10.357 \pm 0.004$ \\
2MASS 08262257+2659457                            & $B-V = +0.475 \pm 0.005$ \\
$\alpha = 08^{\rm h} 26^{\rm m} 22\fs 6$  (J2000) & $V-R_{\rm C} = +0.278 \pm 0.008$ \\
$\delta = +26\arcdeg 59\arcmin 46\arcsec$ (J2000) & $V-I_{\rm C}\, = +0.530 \pm 0.006$ \\
Period = $3\fd35002 \pm 0.00002$       &         $J\, = \: \phn 9.478 \pm 0.021$ \\
Depth = $0.014 \pm 0.002$ mag          &         $H\, = \: \phn 9.286 \pm 0.022$ \\
                                       & $K_{\rm s} = \: \phn 9.207 \pm 0.021$ \\
\enddata
\end{deluxetable}

\begin{deluxetable}{llll}
\tablewidth{0pt}
\tablecaption{HIRES spectral standards \label{t:hires_std}}
\tablehead{
\colhead{Star} & \colhead{Spectral} & \colhead{Radial} &
\colhead{Spectral Type/Radial Velocity} \\
\colhead{} & \colhead{Type} & \colhead{Velocity} &
\colhead{References}}
\startdata
HD 78154A &  F6IV   & $-2.90 \pm 0.39$ & Simbad/\citet{deme99} \\
HD 84737  &  G0.5Va & $+4.90 \pm 0.3$  & \citet{Keen89}/\citet{Nide02} \\
HD 10151  &  G8V    & $-5.65 \pm 0.3$  & \citet{Keen89}/\citet{Nide02} \\
\enddata
\end{deluxetable}

\begin{deluxetable}{lcccccccc}
\tablewidth{0pt}
\tablecaption{Properties of the stars in \GSC\ composing the blend. 
\label{t:blendfit}}
\tablehead{
\colhead{} & \colhead{Mass} & \colhead{$T_{\rm eff}$} &
\colhead{Radius} & \colhead{$\log g$} & \colhead{$M_V$} &
\colhead{$M_{R_{\rm C}}$} & \colhead{$V$} & \colhead{$R_{\rm C}$} \\
\colhead{~~~~~~~~Star~~~~~~~~~} & \colhead{($M_{\sun}$)} &
\colhead{(K)} & \colhead{($R_{\sun}$)} & \colhead{(cgs)} &
\colhead{(mag)} & \colhead{(mag)} & \colhead{(mag)} &\colhead{(mag)}}
\startdata
F star \dotfill & 1.68 & 6497 & 2.74 & 3.79 & \phn2.09 & 1.81 & 10.49 & 10.21 \\
G star \dotfill & 1.12 & 6055 & 1.11 & 4.40 & \phn4.38 & 4.05 & 12.78 & 12.45 \\
M star \dotfill & 0.36 & 3721 & 0.36 & 4.89 & 10.29 & 9.36 & 18.69 & 17.76 \\
\enddata
\end{deluxetable}

\begin{deluxetable}{ccccc}
\tablewidth{0pt}
\tabletypesize{\scriptsize}
\tablecaption{Radial velocity measurements of \GSC\ (heliocentric frame) measured 
with TODCOR. \label{t:todcor}}
\tablehead{
\colhead{HJD} & \colhead{RV$_{\rm F}$} & \colhead{RV$_{\rm EB}$} &
\colhead{(O$-$C)$_{\rm EB}$}  & \colhead{Orbital} \\
\colhead{\hbox{~~(2,400,000$+$)~~}} & \colhead{(\kms)} &
\colhead{(\kms)} & \colhead{(\kms)} & \colhead{Phase\tablenotemark{a}}}
\startdata
    53041.7973\dotfill &  $+$34.13 &  \phn$+$7.36 &    $-$5.41  &   0.102 \\
    53042.8578\dotfill &  $+$35.30 & $+$12.12 &    $-$4.95  &   0.419 \\
    53043.8144\dotfill &  $+$36.14 & $+$74.83 &    $+$2.42  &   0.705 \\
    53045.8283\dotfill &  $+$34.62 &  \phn$+$0.32 &    $+$0.53  &   0.306 \\
    53046.6734\dotfill &  $+$36.20 & $+$44.89 &    $-$4.44  &   0.558 \\
    53046.6880\dotfill &  $+$34.48 & $+$50.91 &    $+$0.61  &   0.562 \\
    53046.8362\dotfill &  $+$35.52 & $+$61.10 &    $+$1.66  &   0.607 \\
    53047.6597\dotfill &  $+$35.38 & $+$62.57 &    $-$3.75  &   0.852 \\
    53047.8879\dotfill &  $+$35.22 & $+$54.27 &    $+$0.23  &   0.920 \\
    53048.6637\dotfill &  $+$36.07 &  \phn$+$3.65 &    $-$0.84  &   0.152 \\
    53049.7189\dotfill &  $+$36.87 & $+$24.00 &    $-$3.85  &   0.467 \\
    53050.4997\dotfill &  $+$35.53 & $+$72.08 &    $-$0.01  &   0.700 \\
    53052.6484\dotfill &  $+$35.18 &  \phn$+$4.98 &    $+$1.36  &   0.342 \\
    53054.6574\dotfill &  $+$34.50 & $+$53.23 &    $+$3.71  &   0.941 \\
    53055.7094\dotfill &  $+$35.18 &  \phn$+$1.54 &    $+$4.05  &   0.255 \\
    53058.6548\dotfill &  $+$37.28 & $+$10.72 &    $+$3.61  &   0.134 \\
    53059.5737\dotfill &  $+$36.76 & $+$13.34 &    $-$1.63  &   0.409 \\
    53060.7572\dotfill &  $+$34.96 & $+$73.56 &    $-$0.29  &   0.762 \\
    53061.6383\dotfill &  $+$35.22 & $+$27.17 &    $-$2.55  &   0.025 \\
    53062.6745\dotfill &  $+$35.45 &  \phn$+$2.23 &    $-$0.49  &   0.334 \\
    53070.7608\dotfill &  $+$35.51 & $+$70.86 &    $-$3.10  &   0.748 \\
    53071.7702\dotfill &  $+$35.62 & $+$24.18 &    $+$0.16  &   0.049 \\
    53072.7035\tablenotemark{b}\dotfill &  $+$34.79 &  \phn$-$6.24 &    $-$8.22  &   0.328 \\
    53073.7394\dotfill &  $+$35.10 & $+$67.39 &    $+$2.63  &   0.637 \\
    53074.7501\dotfill &  $+$34.73 & $+$52.63 &    $+$2.61  &   0.939 \\
    53075.6526\dotfill &  $+$35.19 &  \phn$+$0.74 &    $+$1.97  &   0.208 \\
    53076.8331\dotfill &  $+$35.46 & $+$49.83 &    $-$0.14  &   0.561 \\
    53127.6627\dotfill &  $+$34.44 & $+$72.37 &    $-$1.39  &   0.734 \\
    53128.7340\dotfill &  $+$36.01 & $+$22.02 &    $-$1.08  &   0.053 \\
    53129.6589\dotfill &  $+$36.73 &  \phn$+$7.14 &    $+$4.99  &   0.330 \\
    53130.7108\dotfill &  $+$34.61 & $+$71.89 &    $+$6.17  &   0.644 \\
    53132.6602\dotfill &  $+$34.89 &  \phn$+$2.60 &    $+$4.68  &   0.225 \\
    53135.6883\dotfill &  $+$35.11 &  \phn$+$8.70 &    $+$0.76  &   0.129 \\
\enddata
\tablenotetext{a}{Referred to the ephemeris of Eq.[1].}
\tablenotetext{b}{The observation taken on this date shows the largest
velocity residual, and happens to be contaminated by moonlight. It is
the same observation omitted from Figure~\ref{f:bisectors}.}
\end{deluxetable}

\begin{deluxetable}{lccc}
\tablewidth{0pt}
\tablecaption{Predicted total apparent magnitudes from our blend model for \GSC,
compared with the measured brightness in the optical and near infrared. 
\label{t:mags}}
\tablehead{
\colhead{} & \colhead{Predicted} & \colhead{Observed} & \colhead{O$-$C} \\
\colhead{} & \colhead{brightness\tablenotemark{a}} &
\colhead{brightness} & \colhead{difference} \\
\colhead{Passband} & \colhead{(mag)} & \colhead{(mag)} & \colhead{(mag)} }
\startdata
$B$\dotfill & 10.844 & 10.832 & $-$0.012 \\
$V$\dotfill & 10.364 & 10.357 & $-$0.007 \\
$R_{\rm C}$\dotfill & 10.079 & 10.079 & \phantom{$-$}0.000 \\
$I_{\rm C}$\dotfill & \phn9.807 & \phn9.827 & $+$0.020 \\
$J$\dotfill & \phn9.462 & \phn9.478 & $+$0.016 \\
$H$\dotfill & \phn9.285 & \phn9.286 & $+$0.001 \\
$K_{\rm s}$\dotfill & \phn9.217 & \phn9.207 & $-$0.010 \\
\enddata
\tablenotetext{a}{The $JHK$ magnitudes from the \cite{Gira00}
models have been converted to the 2MASS system for comparison with the
observed quantities, following \cite{Carp01}.}
\end{deluxetable}

\end{document}